\documentstyle[prl,aps,psfig]{revtex}

\draft

\begin{document}
\twocolumn[\hsize\textwidth\columnwidth\hsize\csname
@twocolumnfalse\endcsname

\draft

\title{
\hbox to\hsize{\large Submitted to Phys.~Rev.~Lett.\hfil E-Print
astro-ph/9604093}
\vskip1.55cm
Highest Energy Cosmic Rays, Grand Unified Theories, 
and the Diffuse Gamma-Ray Background}

\author{G\"unter Sigl and Sangjin Lee}
\address{Department of Astronomy \& Astrophysics,
Enrico Fermi Institute, The University of Chicago, Chicago, IL~~60637-1433\\
and NASA/Fermilab Astrophysics Center,
Fermi National Accelerator Laboratory, Batavia, IL~~60510-0500}

\author{Paolo Coppi}
\address{Department of Astronomy,
Yale University, New Haven, CT 06520-8101}

\maketitle

\begin{abstract}
We explore scenarios where the highest energy cosmic rays (HECR)
are produced by new particle physics near the grand unification
scale. Using detailed numerical
simulations of extragalactic cosmic and $\gamma$-ray
propagation, we show the existence of a significant
parameter space for which such scenarios are consistent with
all observational constraints. An average fraction of
$\simeq10\%$ $\gamma$-rays in the total cosmic ray flux around
$10\,$EeV ($10^{19}\,$eV) would imply both a non-acceleration
origin of HECR and a large scale extragalactic magnetic field
$\lesssim10^{-11}\,$G. Proposed observatories for
ultra-high energy cosmic rays should be able to test for
this signature.

\end{abstract}

\vskip2pc]

\narrowtext

The HECR events observed above
$100\,$EeV~\cite{fe1,agasa1} are difficult to explain within
conventional models involving first order Fermi
acceleration of charged particles at astrophysical
shocks~\cite{Blandford}. It is hard to accelerate 
protons and heavy nuclei up to such energies even in the most
powerful astrophysical objects~\cite{Hillas}, like radio
galaxies and active galactic nuclei. Also, nucleons above
$\simeq70\,$EeV undergo
photopion production on the cosmic microwave background (CMB),
which is known as the Greisen-Zatsepin-Kuzmin (GZK) effect~\cite{GZK}
and limits the distance to possible sources to less than
$\simeq100\,$Mpc~\cite{SSB}. Heavy nuclei are photodisintegrated
in the CMB within a few Mpc~\cite{Puget}. There are no 
obvious astronomical sources within $\simeq 100$ Mpc of the
Earth.

A way around these difficulties is to suppose the HECR are created
directly at energies comparable to or exceeding the observed ones rather
than being accelerated from lower energies. In the current
versions of such ``top-down'' (TD) scenarios, predominantly
$\gamma$-rays and neutrinos are initially produced at ultra-high
energies (UHEs) by the quantum
mechanical decay of supermassive elementary ``X" particles related
to some grand unified theory (GUT). Such X particles could be
released from topological defect relics of phase
transitions which might have been caused by spontaneous breaking
of GUT symmetries in the early universe~\cite{BHS}. 
TD models of this type are attractive 
because they predict injection spectra
which are considerably harder than shock acceleration spectra
and, unlike the GZK effect for nucleons, there is no threshold
effect in the attenuation of UHE $\gamma$-rays which dominate the
predicted flux.

There has been considerable discussion in the literature whether
the $\gamma$-ray, nucleon, and neutrino fluxes predicted by TD
scenarios are consistent with observational data and constraints
at any energy~\cite{Chi,SJSB,PJ,PS}. The 
absolute flux levels predicted by TD models are in general extremely
uncertain~\cite{GK}. Accordingly, in this letter we treat
the production rate of decaying X particles as a free parameter
to be adjusted to match data and constraints. 
(We note that TD scenarios such as annihilation of magnetic 
monopole-antimonopole 
pairs~\cite{BS} {\it can} yield HECR fluxes consistent with observations
without violating bounds on monopole abundances.)
Under this ``optimal'' assumption, we then use 
detailed numerical simulations of
extragalactic cosmic and $\gamma$-ray propagation to
show that TD models are 
still viable for an interesting range of parameters.
We also explore a signature for TD
mechanisms based on the isotropic component of the $\gamma$-ray
flux below $100\,$EeV, particularly around $10\,$EeV.
The exposure required to test this signature is significantly smaller than for
measurements above $100\,$EeV (e.g., as proposed in
Refs.~\cite{SLSB} and~\cite{ABS}),  as long as discrimination
between $\gamma$-rays and charged cosmic rays (CRs) is possible at a
level of a few percent. This is within the reach of
proposed experiments~\cite{Cronin}.

{\it Top-Down Models.}
The X particles released, say,
in the annihilation or collapse of defects such as cosmic
strings, monopoles, or domain walls
could be gauge bosons, Higgs bosons, superheavy fermions,
etc.~depending on the specific GUT. Thses X particles would have
a mass $m_X$ comparable to the symmetry breaking scale and would
rapidly decay typically into a lepton and a quark of roughly
comparable energy. We take the primary lepton produced in
a decay to be an electron with energy $m_X/2$. (Prior calculations
have ignored this lepton which is not a good approximation given
the lepton's energy.) The quark interacts strongly and 
hadronizes into nucleons ($N$s) and pions, the latter 
decaying in turn into $\gamma$-rays, electrons, and neutrinos. 
Given the X particle production rate, $dn_X/dt$, the effective
injection spectrum of particle species $a$ ($a=\gamma,N,e^\pm,\nu$) 
via the hadronic channel can be
written as $(dn_X/dt)(2/m_X)(dN_a/dx)$,
where $x \equiv 2E/m_X$, and $dN_a/dx$ is the relevant
fragmentation function. 
For the total hadronic fragmentation function $dN_h/dx$ we use
solutions of the QCD evolution equations in modified leading
logarithmic approximation which provide good fits to accelerator
data at LEP energies~\cite{detal}.
We assume that about 3\% of the total hadronic content consists of
nucleons and the rest is produced as pions and distributed equally among
the three charge states. The standard pion decay spectra then
give the injection spectra of $\gamma$-rays, electrons, and
neutrinos. The X particle injection rate is assumed to be
spatially uniform and in the matter-dominated era can be
parametrized as $dn_X/dt\propto t^{-4+p}$~\cite{BHS},
where $p$ depends on the specific defect scenario. In this
letter we focus on the case $p=1$ which is representative for a
network of ordinary cosmic strings~\cite{BR} and
annihilation of monopole-antimonopole pairs~\cite{BS}.

{\it Numerical Simulations.}
The $\gamma$-rays and electrons produced by  X particle decay
initiate  electromagnetic
(EM) cascades on low energy radiation fields such as the
CMB. The high energy photons undergo electron-positron pair
production (PP; $\gamma \gamma_b \rightarrow e^- e^+$), and 
at energies below $\sim 10^{14}$ eV they interact mainly with 
the universal infrared and optical (IR/O) background, while above 
$\sim 100$ EeV  they interact mainly with the universal radio background (URB).
In the Klein-Nishina regime, where the center of mass energy is
large compared to the electron mass, one of the outgoing particles usually
carries most of the initial energy. This ``leading''
electron (positron) in turn can transfer almost all of its energy to
a background photon via inverse
Compton scattering (ICS; $e \gamma_b \rightarrow e' \gamma$).
EM cascades are driven by this cycle of PP and ICS.
The energy degradation of the ``leading'' particle in this cycle
is slow, whereas the total number of particles grows
exponentially with time. This makes a standard Monte Carlo
treatment difficult. We have therefore used an 
an implicit numerical scheme to solve the relevant kinetic 
equations. A detailed account of our transport equation approach
is in Ref.~\cite{Lee}. 
We include all
EM interactions that influence the $\gamma$-ray spectrum in the energy range
$10^8\,{\rm eV} < E < 10^{25}\,$eV, namely PP, ICS, triplet pair
production (TPP; $e \gamma_b
\rightarrow e e^- e^+$), and double pair production ($\gamma \gamma_b
\rightarrow e^-e^+e^-e^+$).
The relevant nucleon interactions implemented are
pair production by protons ($p\gamma_b\rightarrow p e^- e^+$),
photoproduction of single or multiple pions ($N\gamma_b \rightarrow N
\;n\pi$, $n\geq1$), and neutron decay. Production of secondary
$\gamma$-rays, electrons, and neutrinos by pion decay is also included.
We assume a flat universe with no cosmological constant,
and a Hubble constant of $H_0=75\,{\rm km}\;{\rm sec}^{-1}{\rm
Mpc}^{-1}$ throughout.
An important difference with respect to past work is that we follow
{\it all} produced particles,
whereas the often-used continuous energy loss (CEL)
approximation (e.g., \cite{ABS}) follows
only the leading cascade particles. We find that the CEL approximation
can significantly underestimate the cascade flux at lower energies.

Similar studies using somewhat different numerical techniques
have been performed for the case of discrete sources
injecting $\gamma$-rays and nucleons monoenergetically~\cite{PJ}
and more recently for fragmentation functions $\propto
x^{-1.5}(1-x)^2$ and spatially uniform injection~\cite{PS}.

\begin{figure}[ht]
\centerline{\psfig{file=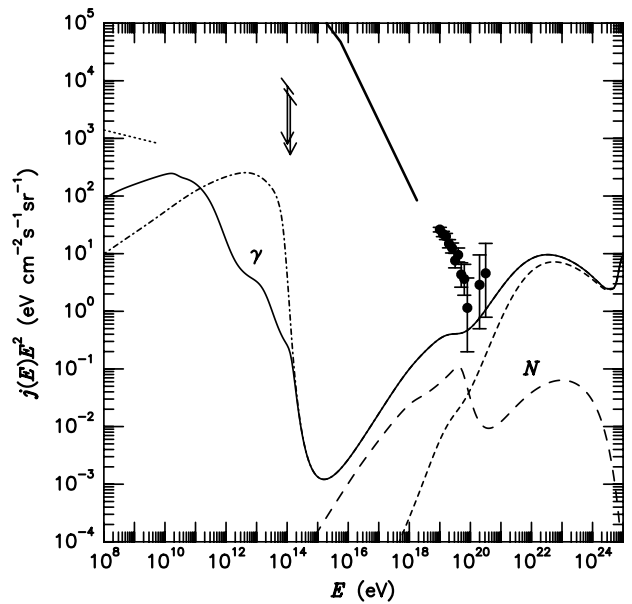,width=3.2in}}
\medskip
\caption[...]{Predictions for the differential fluxes of $\gamma$-rays (solid
line) and nucleons (long dashed line) by a TD model
characterized by $p=1$, $m_X = 2\times10^{16}\,$GeV, for no EGMF.
The dashed line shows the $\gamma$-ray flux predicted by the
CEL approximation. The dash-dotted line shows the result when there
is no IR/O background. Also shown are the combined data from the Fly's
Eye~\cite{fe1} and the AGASA~\cite{agasa1} experiments above $10\,$EeV
(dots with error bars), piecewise power law fits to the observed
charged CR flux (thick
solid line) and experimental upper limits on the $\gamma$-ray
flux at $1-10\,$GeV from EGRET data~\cite{CDKF} (dotted
line on left margin). The arrows indicate limits on the
$\gamma$-ray flux from Ref.~\cite{hegra}.
\label{F1}}
\end{figure}

\begin{figure}[ht]
\centerline{\psfig{file=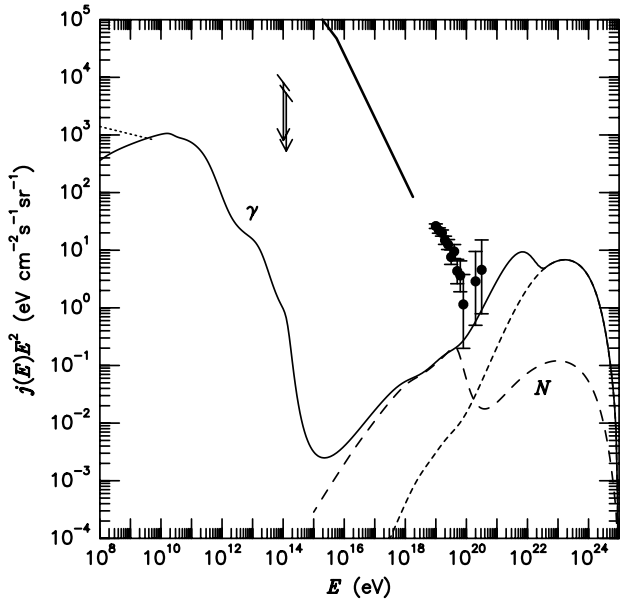,width=3.2in}}
\medskip
\caption[...]{Same as Fig.~\ref{F1}, but for an EGMF of
$10^{-9}\,$G.
\label{F2}}
\end{figure}

{\it Results.}
Fig.~\ref{F1} shows the results for the $\gamma$-ray and nucleon
fluxes from a typical TD scenario, assuming no
EGMF, along with current observational constraints on
the $\gamma$-ray flux. The spectrum was normalized in the best
possible way to allow for an explanation of the observed HECR
events, assuming their consistency with a nucleon or
$\gamma$-ray primary (although a $\gamma$-ray primary is
somewhat disfavored~\cite{HVSV}). The flux below $\lesssim 20$ EeV
is presumably due to conventional acceleration and was not fit.
The shapes of
our spectra are similar to those of Ref.~\cite{PS}. However,
they normalize their spectra to match the
observed differential flux at $300\,$EeV,  which then 
leads to an overproduction of the
integral flux at higher energies. We remark that above $100\,$
EeV, the fits shown in Figs.~\ref{F1} and~\ref{F2} have
likelihood significances above 50\% (see Ref.~\cite{SLSB} for
details) and are consistent with the integral flux above
$300\,$EeV esitmated in Refs.~\cite{fe1,agasa1}.
Since the energy injected at high
redshifts is recycled by cascading to lower energies, TD models
are significantly constrained~\cite{Chi,SJSB} by current
limits on the diffuse $\gamma$-ray background at
$1-10\,{\rm GeV}$~\cite{CDKF}.
Note that the IR/O background strongly depletes the background
in the range $10^{11} - 10^{14}\,$eV, recycling it to
energies below $10\,$GeV (see Fig.~\ref{F1}).
The predicted background is {\it not} very sensitive to
the specific IR/O background model, however~\cite{AC}.
Constraints from limits on CMB distortions and light
element abundances from $^4$He-photodisintegration are
comparable to the bound from the directly observed
$\gamma$-rays~\cite{SJSB}. The scenario in Fig.~\ref{F1}
obeys all current constraints within the normalization
ambiguities and is therefore quite viable.

Fig.~\ref{F2} shows results for the same TD scenario as in
Fig.~\ref{F1}, but for a high EGMF $\sim 10^{-9}$ G
(near the current upper
limit~\cite{Kronberg}). 
In this case, rapid synchrotron cooling of the initial cascade pairs quickly
transfers energy out of the UHE range. The UHE $\gamma-$ray flux then depends
mainly on the absorption length due to pair production and is typically
much lower~\cite{ABS,LOS}. (Note, though, that for $m_X\gtrsim
10^{25}$ eV, the synchrotron radiation from these pairs
can be above 100 EeV, and the UHE flux
is then not as low as one might expect.) To match the HECR flux, we must inject
more X particles, which leads to a factor $\sim$ 5 increase in the 
$\gamma-$ray background expected below 10 GeV. For a high EGMF, 
the constraints from the flux limits below $10\,$GeV are much more severe 
and in general, TD scenarios in a high EGMF are only marginally allowed. 

The energy loss and absorption lengths for UHE nucleons and photons
are short ($\lesssim 100$ Mpc). Thus, their predicted UHE fluxes are
independent of cosmological evolution. The $\gamma-$ray flux
below $\simeq 10^{11}\,$eV, however, scales as the
total X particle energy release integrated over all redshifts
and increases with decreasing $p$~\cite{SJSB}. For
$m_X=2\times10^{16}\,$GeV,
scenarios with $p<1$ are therefore ruled
out (see Figs.~\ref{F1} and ~\ref{F2}), whereas
constant comoving injection rates ($p=2$) are well within the
limits. As the EM flux above $\simeq10^{22}\,$eV is
efficiently recycled to lower energies, the constraint on $p$
is insensitive to $m_X$ for low EGMF. This in contrast to
earlier CEL-based analytical estimates
~\cite{Chi,SJSB}.

\begin{figure}[ht]
\centerline{\psfig{file=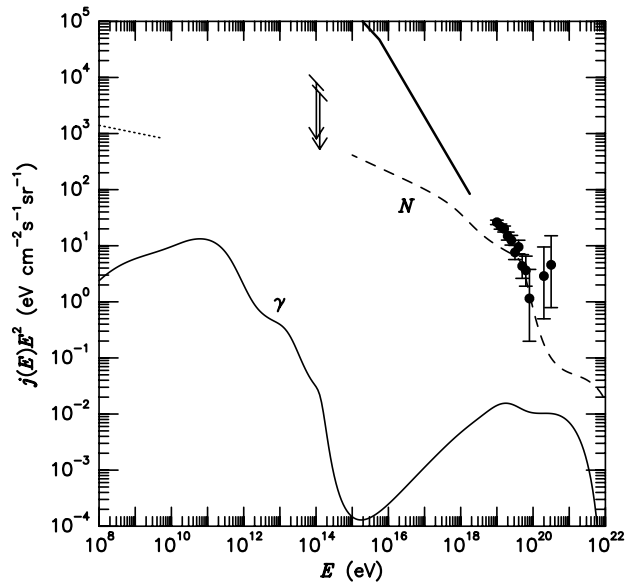,width=3.2in}}
\medskip
\caption[...]{Predictions for the differential fluxes of $\gamma$-rays (solid
line) and nucleons above $10^{15}\,$eV (long dashed line) by a
uniform, constant comoving density of shock
acceleration sources up to a redshift of 4, injecting protons with a spectrum
$\propto E^{-2.3}$ up to $10^{22}\,$eV, for a vanishing EGMF.
\label{F3}}
\end{figure}

We now turn to signatures of TD models at UHEs. For low EGMF
(e.g., Fig.~\ref{F1}), the full cascade calculation predicts
$\gamma$-ray fluxes below $100\,$EeV an 
order of magnitude higher than those obtained
using the CEL approximation. Again, this shows the importance
of non-leading particles in the development of unsaturated EM
cascades at energies below $\sim10^{22}\,$eV.
Our numerical simulations give a $\gamma$/CR flux ratio at $10\,$EeV of
$\simeq0.1$. The experimental exposure
required to detect a $\gamma$-ray flux at that level is
$\simeq4\times10^{19}\,{\rm cm^2}\,{\rm sec}\,{\rm sr}$, about a
factor 10 smaller than the current total experimental exposure.
These exposures are well within reach of the proposed
Pierre Auger Cosmic Ray Observatories~\cite{Cronin}, which may be able to
detect a neutral CR component down to a level of 1\% of the total
flux.  In contrast, if the EGMF exceeds $\sim 10^{-11}\,$G, then UHE
cascading is inhibited, resulting in a lower UHE 
$\gamma$-ray spectrum. In the $10^{-9}$ G scenario of Fig.~\ref{F2},
the $\gamma$/CR flux ratio at $10\,$EeV is $\simeq 4 \times
10^{-3}$, significantly lower than for no EGMF.

Another not well known factor affecting UHE
$\gamma$-ray propagation is the URB for which we used the
spectrum suggested in Ref.~\cite{Clark}. A higher
overall URB amplitude correspondingly reduces the
$\gamma$-ray flux, but without significantly changing its
spectral shape. Thus, as
long as $\gamma$-rays dominate nucleons in the TD
component above $\simeq100\,$EeV and the total flux is
normalized to the HECR events,
predictions for the $\gamma$/CR flux ratio below
$\simeq100\,$EeV are essentially independent of the URB
amplitude and the hadronic fraction at injection.
An URB cutoff frequency lower than $2\,$MHz (the value we took)
affects this ratio in a less trivial way with a
tendency to smaller $\gamma$/CR values.

Fig.~\ref{F3} shows spectra resulting from a 
typical non-TD scenario, in this case a uniform
distribution of shock acceleration sources.
Such a scenario gives a UHE CR spectrum with a
GZK cutoff and $\gamma$-rays are only produced as
secondaries. Our treatment
of multiple pion production by nucleons leads to
secondary $\gamma$-ray fluxes somewhat higher than in
Refs.~\cite{PJ,YT}. The key point is that
the (isotropic) $\gamma$/CR flux
ratio is $\lesssim 10^{-3}$ at $10\,$EeV, much smaller than
predicted by TD models in a small EGMF. Ratios as high as
$10\%$ can only be reached in the direction of powerful nearby
acceleration sources.
The secondary $\gamma$-ray flux generally decreases still further
with decreasing maximum injection energy and increasing
EGMF~\cite{LOS}.

In summary, some TD-type scenarios for the HECR origin 
are still unconstrained by
current data and bounds on $\gamma$-ray and UHE CR fluxes.
For example, if the mean EGMF is $\lesssim10^{-9}\,$G, spatially
uniform annihilation of magnetic monopoles and antimonopoles is
still a viable model for GUT scales up to $10^{16}\,$GeV.
A solid angle averaged $\gamma$/CR flux ratio  $\simeq 10\%$
at $\sim 10\,$EeV would be hard
to explain by a conventional acceleration origin of HECRs,
and assuming the TD picture holds, would place an independent
upper limit of $\simeq10^{-11}\,$G on
the poorly known EGMF on scales of a few to tens of Mpc. 
Absence of a high $\gamma$/CR flux at this level either rules
out a TD origin for HECRs or implies an 
EGMF strength $\gtrsim10^{-11}\,$G.
A test of this signature should be possible with
currently proposed experiments. TD models also predict
significant neutrino fluxes. Implications of this will be
considered in a separate publication~\cite{SLC}.

We thank W.~Ochs, D.~N.~Schramm, J.~W.~Cronin, Chris Hill, and
F.~A.~Aharonian for stimulating discussions and
Shigeru Yoshida for comments. This work was supported by DOE,
NSF and NASA at the University of Chicago, by the DOE and by NASA through
grant NAG5-2788 at Fermilab, and by the Alexander-von-Humboldt
Foundation. SL
acknowledges the support of the POSCO Scholarship Foundation in Korea.


\begin{references}

\bibitem{fe1} D.~J.~Bird {\it et al.}, Phys.~Rev.~Lett.~{\bf 71}, 3401 (1993);
Astrophys.~J.~{\bf 424}, 491 (1994); {\it ibid.} {\bf 441}, 144 (1995).

\bibitem{agasa1} N.~Hayashida {\it et al.}, Phys.~Rev.~Lett.~{\bf 73}, 3491
(1994); S.~Yoshida {\it et al.}, Astropart.~Phys.~{\bf 3}, 105 (1995).

\bibitem{Blandford} for a review see, e.g., R.~Blandford and
D.~Eichler, Phys. Rep. {\bf 154}, 1 (1987).

\bibitem{Hillas} A.~M.~Hillas, Ann. Rev. Astron. Astrophys. {\bf
22}, 425 (1984).

\bibitem{GZK} K.~Greisen, Phys.~Rev.~Lett.~{\bf 16}, 748 (1966);
G.~T.~Zatsepin and V.~A.~Kuzmin, Pisma Zh.~Eksp.~Teor.~Fiz.~{\bf 4}, 114
(1966) [JETP.~Lett.~{\bf 4}, 78 (1966)].

\bibitem{SSB} G.~Sigl, D.~N.~Schramm, and P.~Bhattacharjee,
Astropart. Phys. {\bf 2}, 401 (1994).

\bibitem{Puget} J.~L.~Puget, F.~W.~Stecker, and J.~H.~Bredekamp,
Astrophys. J. {\bf 205}, 638 (1976).

\bibitem{BHS} P.~Bhattarcharjee, C.~T.~Hill, and D.~N.~Schramm,
Phys.~Rev.~Lett.~{\bf 69}, 567 (1992).

\bibitem{Chi} X.~Chi {\it et al.}, Astropart. Phys. {\bf 1}, 129
(1993); {\it ibid.} {\bf 1}, 239 (1993).

\bibitem{SJSB} G.~Sigl, K.~Jedamzik, D.~N.~Schramm, and V.~Berezinsky,
Phys.~Rev.~D {\bf 52}, 6682 (1995).

\bibitem{PJ} R.~J.~Protheroe and P.~A.~Johnson,
Astropart.~Phys. {\bf 4}, 253 (1996).

\bibitem{PS} R.~J.~Protheroe and T.~Stanev, report ADP-AT-96-6,
astro-ph/9605036, submitted to Phys. Rev. Lett.

\bibitem{GK} A.~J.~Gill and T.~W.~B.~Kibble, Phys.~Rev.~D {\bf
50}, 3660 (1994).

\bibitem{BS} P.~Bhattarcharjee and G.~Sigl, Phys.~Rev.~D {\bf 51}, 4079 (1995).

\bibitem{SLSB} G.~Sigl, S.~Lee, D.~N.~Schramm, and P.~Bhattacharjee, Science
{\bf 270}, 1977 (1995).

\bibitem{ABS} F.~A.~Aharonian, P.~Bhattacharjee, and D.~N.~Schramm,
Phys.~Rev.~D {\bf 46}, 4188 (1992).

\bibitem{Cronin} M.~Boratav {\it et al.}, eds., Nucl. Phys. B
(Proc. Suppl.) {\bf 28B} (1992).

\bibitem{detal} Yu.~L.~Dokshitzer, V.~A.~Khoze, A.~H.~M\"uller,
and S.~I.~Troyan, {\sl Basics of
Perturbative QCD} (Editions Frontieres, Singapore, 1991).

\bibitem{BR} P.~Bhattacharjee and N.~C.~Rana, Phys.~Lett.~B {\bf 246}, 365
(1990).

\bibitem{Lee} S.~Lee, report FERMILAB-Pub-96/066-A,
astro-ph/9604098, submitted to Phys.~Rev.~D.

\bibitem{HVSV} F.~Halzen, R.~A.~Vazquez, T.~Stanev, and
H.~P.~Vankov, Astropart. Phys. {\bf 3}, 151 (1995).

\bibitem{CDKF} A.~Chen, J.~Dwyer, and P.~Kaaret,
Astrophys. J. {\bf 463}, 169 (1996); C.~E.~Fichtel, {\it
Proc. 3rd Compton Observatory Symposium},
Astron. Astrophys. Suppl., in press.

\bibitem{hegra} A.~Karle {\it et al.}, Phys.~Lett.~B {\bf 347}, 161 (1995).

\bibitem{AC} P.~Coppi and F.~Aharonian, submitted to Astrophys.~J.~Lett.

\bibitem{LOS} S.~Lee, A.~V.~Olinto, and G.~Sigl,
Astrophys. J. {\bf 455}, L21 (1995).

\bibitem{Kronberg} P.~P.~Kronberg, Rep. Prog. Phys. {\bf 57},
325 (1994).

\bibitem{Clark} T.~A.~Clark, L.~W.~Brown, and J.~K.~Alexander, Nature {\bf
228}, 847 (1970).

\bibitem{YT} S.~Yoshida and M.~Teshima, Prog.~Theor.~Phys.~{\bf 89}, 833
(1993).

\bibitem{SLC} G.~Sigl, S.~Lee, and P.~Coppi, in preparation.

\end{references}
\end{document}